# Correlation between electrical and magnetic properties of phase separated manganites studied with a General Effective Medium model

J. Sacanell,[a,*] M. Quintero,[a,b] F. Parisi,[a,b] L. Ghivelder,[c] A. G. Leyva,[a,b] P. Levy[a]

[a]*Departamento de Física, Centro Atómico Constituyentes, CNEA, Av. Gral. Paz 1499, (1650) San Martín, Prov. de Buenos Aires, Argentina.*

[b]*Escuela de Ciencia y Tecnología, UNSAM, San Martín, Buenos Aires, Argentina.*

[c]*Instituto de Fisica, Universidade Federal do Rio de Janeiro, Caixa Postal 68528, Rio de Janeiro, RJ 21941-972, Brazil.*

**Abstract**

We have performed electrical resistivity and DC magnetization measurements as a function of temperature, on polycrystalline samples of phase separated $La_{5/8-y}Pr_yCa_{3/8}MnO_3$ (y = 0.3). We have used the General Effective Medium Theory to obtain theoretical resistivity vs. temperature curves corresponding to different fixed ferromagnetic volume fraction values, assuming that the sample is a mixture of typical metallic – like and insulating manganites. By comparing this data with our experimental resistivity curves we have obtained the relative ferromagnetic volume fraction of our sample as a function of temperature. This result matches with the corresponding magnetization data in excellent agreement, showing that a mixed phase scenario is the key element to explain both the magnetic and transport properties in the present compound. © 2006 Elsevier Science. All rights reserved

## 1. Introduction

Intrinsic multiscale coexistence of different phases in complex oxides is nowadays one of the most important topics in the study of strongly correlated electron systems [1]. Among the large family of systems displaying this behavior, mixed valent manganese oxides, also known as manganites, are one of the most widely studied since the discovery of the colossal magnetoresistance (CMR) effect [2]. The competition among ferromagnetic double exchange [3] and antiferromagnetic superexchange interactions, is responsible for the low temperature state of some manganites, formed by the intrinsic coexistence of typically metallic ferromagnetic (FM) and antiferromagnetic charge ordered (CO) phases, a phenomenon known as phase separation (PS) [4].

In the present work we show how the electric and magnetic properties of one particular phase separated manganite are connected by using a general effective medium model [5]. Within this scenario, the relative ferromagnetic volume fraction can be calculated from electrical transport measurements, and compared with that obtained through magnetization measurements.

The system chosen was the widely studied $La_{5/8-y}Pr_yCa_{3/8}MnO_3$ [6, 7] in which PS can be tuned by the substitution of La ions by Pr ions. Particularly we

———
* Corresponding author. Tel.: +5411 6772 7102; fax: +5411 6772 7121; e-mail: sacanell@cnea.gov.ar

have used a sample with y = 0.3, which presents PS in a wide temperature range [6-12].

## 2. Experimental

Bulk polycrystalline samples of $La_{5/8-y}Pr_yCa_{3/8}MnO_3$ with y = 0.3 (LPCMO(0.3)), were synthesized following the liquid mix method. The obtained powder was pressed into bars with the dimension of $5 \times 1 \times 1$ mm$^3$. DC magnetization measurements were performed in a Quantum Design PPMS system. The electrical resistivity was measured in an He closed cycle criogenerator using the standard four probe method.

## 3. Results and discussion

Figure 1 shows the temperature dependence of the electrical resistivity ($\rho$) and magnetization (*M*) of LPCMO(0.3). The system is paramagnetic insulator at room temperature, the insulating CO phase develops at $T_{co} \sim 220$ K, as evidenced by the increase in $\rho$ and the peak in *M*. With further cooling, nucleation of FM droplets occurs below $T_{C1} = 220$ K, which co-exists with the CO phase in the 220 K – 80 K temperature range [13]. In this range, the metallic FM fraction ( *f* ) of the system remains nearly constant as can be inferred by the plateau in *M(T)* and an insulator to metal transition is observed when *f* reaches the percolation threshold. On cooling below $T_{C2} = 100$ K, a long range FM phase is formed. Magnetization measurements were performed under the application of a 200 Oe magnetic field which is enough to align the FM domains but has no effect on the CO phase.

In the first place, as we want to obtain the *f* vs T dependence corresponding to our sample we have looked for an approach to account for the electrical transport properties of mixtures of materials with different electrical characteristics. We have used the General Effective Medium Theory (GEM) proposed by McLachlan [5], which describes the effective resistivity ($\rho_e$) of a binary mixture as a function of the intrisic resistivities of its constituents. In our case we consider that the sample is a mixture of a pure FM system with resistivity $\rho_{FM}$ and a pure CO one with resistivity $\rho_{co}$. The GEM equation is:

$$f\frac{\rho_e^{1/t} - \rho_{FM}^{1/t}}{\rho_e^{1/t} + A_C\rho_{FM}^{1/t}} + (1-f)\frac{\rho_e^{1/t} - \rho_{co}^{1/t}}{\rho_e^{1/t} + A_C\rho_{co}^{1/t}} = 0$$

The value $A_c = 1/f_c - 1$ depends on the critical percolation fraction $f_c$ and *t* is an exponent which depends on the dimension of the system an the shape of the clusters that form the mixture.

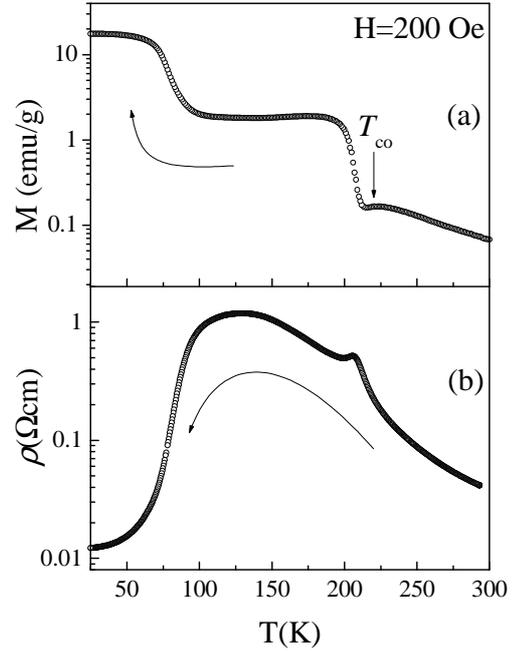

Figure 1: Temperature dependence of (a) *M* (*H*=200 Oe) and (b) $\rho$ for polycrystalline LPCMO(0.3) on cooling. The charge ordering temperature ($T_{co}$) is marked on the graph.

In figure 2 we show the $\rho$ vs *T* dependence for a LPCMO(0.4) sample (thick solid line labeled as *f* = 0) and for a LPCMO(0) sample (thick dashed line labeled as *f* = 1). According to previously reported data, we have assumed that a sample of LPCMO(0.4) is a nearly pure CO system [14], while the LPCMO(0) sample is homogeneously FM [6-8], these curves were used as $\rho_{FM}$ and $\rho_{co}$ as a function of *T*, respectively. We have also added in solid thin lines the theoretical $\rho$ vs *T* for samples with different fixed *f* values, calculated using the GEM equation. The measured $\rho$ vs *T* of the LPCMO(0.3) sample is shown for comparison, from which we can immediately visualize and calculate the temperature dependence of the FM fraction of LPCMO(0.3).

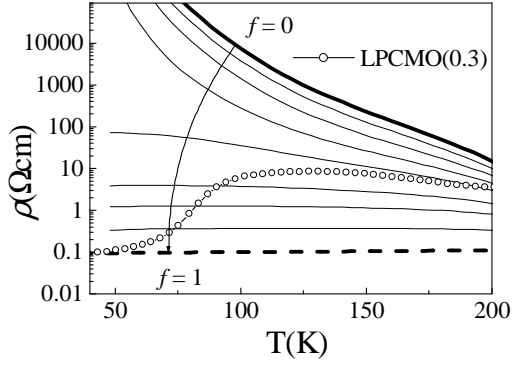

Figure 2: Experimental $\rho$ vs $T$ for samples with $f = 0$ (thick solid line), $f = 1$ (thick dashed line) and for the LPCMO(0.3) sample (circles) on cooling. The curves corresponding to intermediate $f$ values calculated using the GEM equation are shown in thin solid lines.

In figure 3 we show in symbols, the temperature dependence of the FM fraction which was obtained from the magnetization measurement of figure 1 as $f_{exp}(T) = M(T)/M_{FM}$, where $M_{FM}$ is the value of the magnetization for a pure FM sample. We have used the GEM equation and the experimental values of $\rho_{FM}$ and $\rho_{co}$ to calculate $f$ at each temperature. The exponent $t$ was used as fitting parameter, beginning from already reported values previously used for manganites as starting points (see ref. [8, 15, 16]). We have kept $f_c$ near 0.1 because this is the approximated value in which the insulator to metal transition is observed. The best fit of $f_{exp}$ vs $T$ is also shown in the figure, which was attained for $t \sim 2.5$. The graph only shows the temperature range where the system is phase separated.

According to McLachlan [5], $t$ and $f_c$ depend on the shape, the orientation of the constituents and on the dimension of the system. He also stated that an universal value of $t \sim 2$ is expected from 3D percolation theory. Our attempt to fit with this value gave a poor adjustment of the data in the 90 to 200 K range. The value obtained for $t$ is larger than the one expected from universal behavior, but this was already observed in continuous systems [17,18] and also in systems in which tunneling occurs [19]. Then, our result can be related with the fact that the FM fraction is distributed along continuous paths as a kind of "swiss-cheese" with the insulating phase being the randomly placed holes (see ref. [18]). This scenario is also consistent with the magnetic force microscopy images of this compound already reported in [20,21] for $100 \le T \le 120$ K.

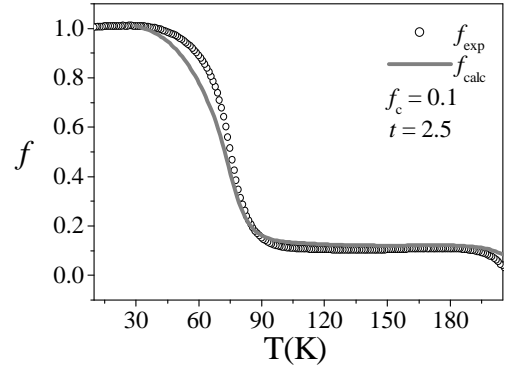

Figure 3: Experimental and calculated temperature dependence for the relative FM fraction ($f$). The obtained fitting parameters are shown on the graph.

## 4. Conclusions

In summary, we have obtained the relative FM volume fraction of the phase separated LPCMO(0.3) manganite as a function of temperature. We have used the GEM equation and the resistivities of the system under study, a fully FM sample and a fully CO sample to perform the calculation. We have observed an excellent agreement between experimental and calculated data. The value obtained for the fitting parameter $t$, even though is larger than the one expected from 3D percolation models, is consistent with a continuum distribution of the FM fraction. The obtained $f_c$ value is rather low as compared to the ones reported in the literature (see [5, 8, 17] and references therein) and thus is a fact that deserves to be further analyzed, however, our confidence in it lays in the fact that it is inferred from the change from insulator to metal in our experimental resistivity curve.

**Acknowledgements**

This work was partially supported by ANPCyT (PICT03-13517) and Fundación Antorchas (Argentina); and CAPES, FAPERJ, and CNPq (Brazil). PL is also a member of CIC CONICET.